\documentclass[pre,aps,twocolumn]{revtex4-1}
\usepackage{graphicx}
\usepackage{amsmath}
\usepackage{amssymb}
\usepackage{color}
\usepackage{natbib}
\usepackage{epsfig}
\def\strutdepth{\dp\strutbox}
\def\nw#1{\strut\vadjust{\kern-\strutdepth\vtop to0pt{\vss\hbox to\hsize
{\hskip\hsize\hskip5pt$\leftarrow$\hss\strut}}}{\em #1}}

\def\We{{\rm We}}

\begin{document}

\title{The effect of surface wettability on inertial pouring flows}

\author{Wilco~Bouwhuis$^{1}$ and Jacco~H.~Snoeijer$^{1,2}$}
\affiliation{$^{1}$Physics of Fluids Group, Faculty of Science and Technology, University of Twente, 7500 AE Enschede, The Netherlands, \\
$^{2}$Mesoscopic Transport Phenomena, Eindhoven University of Technology, 5612 AZ Eindhoven, The Netherlands }

\date{\today}

\begin{abstract}

A liquid poured from a curved solid surface can separate as a steady jet or sheet, or trickle down along the solid surface. It was shown by Duez~{\it et~al.} [Phys.~Rev.~Lett.~{\bf 104},~084503~(2010)] that surface wettability controls the separation of an inertial flow from a solid surface to an unexpected degree, which was further motivated by an inertial-capillary adhesion model. 
In this paper we extend the analysis by a control volume calculation that takes into account the velocity profile within the flowing layer, supported by Boundary Integral potential flow simulations, and the detailed capillary forces induced by the local curvatures of the sheet. Our analysis captures the appearance of a critical Weber number below which no steady separated solutions can be sustained. We investigate the dependence of the critical Weber number on the wettability and sharpness of the edge of the curved solid, and recover the key experimental trends. 
\end{abstract}

\maketitle

\section{Introduction} \label{sec:Introduction}

\begin{figure}[htp!]
\centering
\includegraphics[width=8.0cm]{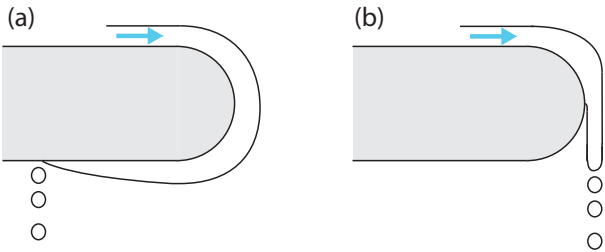}
\caption{Problem sketch. (a) Water flow trickling around the curved (hydrophilic) solid surface, known as the `teapot effect'; (b) The water repellence of the solid (or the flow speed of the liquid) overcomes the trickling behavior.}
\label{Teapot-effect}
\end{figure}

The so-called `teapot effect' is a daily life phenomenon that will be recognized by everyone. When a liquid is poured too slowly from the (hydrophilic) nose of a teapot, a bottle or a beaker, the liquid has the tendency to run down along the underside of the spout,  as depicted in Fig.~\ref{Teapot-effect}a. This remarkable effect already received attention in the 1950s~\cite{Reiner,Keller}, and was explained by Keller and Vanden-Broeck as a purely hydrodynamic (Bernoulli) principle~\cite{Keller,Vanden-Broeck1986,Vanden-Broeck1989}. Since there is a difference in flow velocity between the top and bottom of the liquid film, the pressure is lowest directly above the spout lip; sufficiently low that the flow is pushed further down along the convex solid. In this situation the liquid film completely `trickles' around the solid, and exact potential flow solutions were found in which the edge of the solid was treated as perfectly sharp \cite{Vanden-Broeck1986,Vanden-Broeck1989}. After trickling, a possible separation from the solid could be induced by gravity. 

However, one can overcome the trickling scenario by increasing the flow speed, such that the liquid jet or sheet separates from the solid with an angle, as is sketched in Fig.~\ref{Teapot-effect}b. Simulations of pouring flows have revealed complicated dynamics, particularly when the Reynolds number of the flow becomes of order unity~\citep{Yoon,Kistler}. Moreover, Duez~{\it et~al.}~\citep{Duez} found that surface wettability controls the flow separation to an unexpected degree, also at large Reynolds numbers. This was also confirmed by recent experiments that aim to develop controlled overflow by surface manipulation~\citep{Dong}. Since wettability is determined on a microscopic scale, this highlights the importance of multiple length scales in the problem of pouring flows. In some sense, the experiments and analysis by Duez~{\it et~al.} have revealed a  relatively simple framework for understanding the various regimes of pouring flow. The experiments were performed using an axisymmetric set-up that is the top part of the `liquid bell' geometry~\citep{Clanet,Lhuissier,Dressaire}, for which the relevant parameters are the flow speed, the wettability of the solid, and the radius of curvature at the edge of the impacter. The transition between inertia-induced sheet separation and trickling appeared at well-defined Weber numbers, whose critical values depend on the contact angle and sharpness of the solid edge. Using a scaling argument for the horizontal momentum balance, Duez~{\it et~al.} predicted a dependency on the flow speed of the liquid and the wettability and sharpness of the solid surface, capturing the main experimental trends -- though surprisingly, the experimental trends by Dong~{\it et~al.}~\citep{Dong} suggest a different dependence on the curvature of the edge (linear instead of quadratic). The modeling approach has so far been limited to a force balance in the horizontal (not in the vertical) direction and it has remained a challenge to explicitly capture the origin of the trickling transition. In addition, it is not clear how these results relate to the analytical potential flow solutions~\citep{Vanden-Broeck1986,Vanden-Broeck1989}, which excluded the effect of surface tension and finite edge-curvature.

In this paper, we will perform a control volume analysis for inertial pouring flows over a solid edge of finite curvature, by taking into account the capillary forces induced by the shape of the meniscus and the velocity profile within the liquid. By releasing some geometric constraints imposed in~\cite{Duez}, this will allow for a force balance in the horizontal \emph{and} vertical direction. The model is solved for varying flow parameters and geometric parameters, and we show how this indeed leads to the appearance of a critical Weber number for the trickling transition. In Sec.~\ref{sec:Model}, we will introduce the relevant parameters in the problem and the basic assumptions. This results into a set of coupled equations that provide a prediction for the separation angle. Our main findings will be presented in Sec.~\ref{sec:Results}, where we give both numerical and asymptotic predictions for the critical Weber number. The results will be summarized and compared to experiments in Sec.~\ref{sec:Discussion}.

\section{Model} \label{sec:Model}

\subsection{Definitions and assumptions}

\begin{figure*}[htb!]
\centering
\includegraphics[width=14.0cm]{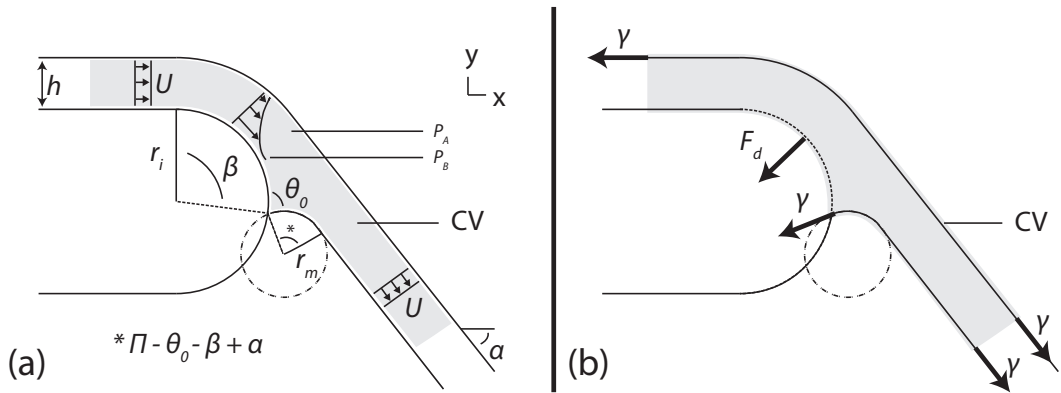}
\caption{Flow around the edge of a solid and two different definitions of the Control Volume (CV), indicated by the gray area. (a) Definition sketch and the Control Surface (CS) located just \emph{inside} the liquid. A liquid film with thickness $h$, which has a uniform flow velocity $U$, bends around a solid surface with radius of curvature $r_i$. $\theta_0$ is the solid-liquid contact angle. $\alpha$ is the global deflection angle of the separated liquid sheet with respect to the horizontal (${\bf x}$) direction, and $\beta$ is the angular width of the wetted fraction of the solid. $\alpha$ and $\beta$ are a priori independent. $r_m$ is the radius of curvature of the circle formed by the meniscus. The figure is not drawn to scale, typically $r_m\ll r_i$. The value of the marked angle $*$ is $\pi\!-\!\theta_0\!-\!\beta\!+\!\alpha$. (b) The force balance and the CS located just \emph{outside} the liquid. The forces, indicated by the arrows, respectively are the capillary forces acting in the interface (denoted by $\gamma$), and the resultant force $F_d$ induced by the pressure difference over the sheet ($P_A\!-\!P_B$).}
\label{Modifiedsetup}
\end{figure*}

A sketch of the problem and the relevant parameters is given in Fig.~\ref{Modifiedsetup}a. Like Ref.~\cite{Duez}, we treat the flow as two-dimensional and define a horizontal $\bf{x}$ direction, and a vertical $\bf{y}$ direction. We assume a steady, laminar, irrotational flow with high Reynolds number. In that case the flow can be considered uniform both at the inflow above the solid and within the separated sheet, with velocity $U$ and the film thickness is $h$. The edge of the solid has a circular shape, characterized by the radius of curvature $r_i$. We assume the sheet separates at an angle $\alpha$ with respect to the horizontal direction, and an important part of the analysis is to determine this angle. The sheet separates from the solid at the position angle $\beta$ defined with respect to the vertical axis. Note that, in general, $\alpha\neq\beta$ and we treat these angles as independent parameters. At the separation point, a small capillary meniscus is formed, which has a radius of curvature $r_m$, which is set by the Young-Laplace pressure difference over the free surface~\cite{Orr,Kralchevsky}. 
Locally, the circle formed by the meniscus crosses the $r_i$-circle with the contact angle $\theta_0$, which is how the surface wettability enters the analysis. Note that $r_m$ is typically much smaller than $r_i$ (Fig.~\ref{Modifiedsetup} not drawn to scale). 

Assuming inviscid flow, the introduced parameters can be expressed by three dimensionless control parameters, which are the Weber number $\We=\rho U^2 h/\gamma$, where $\rho$ and $\gamma$ are the density 
and liquid-gas surface tension 
of the liquid, respectively, the ratio $\tilde{r}_i=r_i/h$, and the contact angle $\theta_0$. For characteristic values around $U \sim 1~\mathrm{m/s}$ and $h\sim ~0.1-1~\mathrm{mm}$, the Weber number for water is of order 1 to 10. The corresponding Reynolds numbers are about $10^2-10^3$, so we can indeed assume laminar flow and neglect viscosity.

Clearly, the formulation presented here closely follows Duez~{\it et~al.}~\cite{Duez}, though some notable differences appear. Importantly, we release the geometric constraint that the tangent of the sheet (with angle $\alpha$ with respect to the vertical direction), is also a tangent line to the edge of the solid (i.e. the circle of radius $r_i$). In this manner we can treat $\alpha$ and $\beta$ as two independent parameters, and the momentum balance can be maintained in both $x$ and $y$ directions. Other minor differences appear below, when estimating the forces acting on the control volume.

We will solve the deflection angle $\alpha$ of the liquid sheet as the result of the hydrodynamic and capillary forces using linear momentum conservation. The mechanism responsible for the liquid sheet bending is that the velocity profile over the curved solid is not uniform, contrarily to the inflow and the outflow of the control volume indicated in Fig.~\ref{Modifiedsetup}. This velocity profile implies a pressure difference across the film, which induces a hydrodynamic force, $F_d$, exerted by the solid on the liquid, pulling the liquid along the curved edge. This is sometimes referred to as the Coanda effect~\cite{Guyon}. The resultant capillary force in the small meniscus is denoted as $F_{men}$, while the effect of surface tension over the top of the curved sheet yields $F_{top}$.  Importantly, we can define the CV in two ways, indicated by Fig.~\ref{Modifiedsetup}a and b, respectively. In Fig.~\ref{Modifiedsetup}a, the Control Surface (CS), is located just \emph{inside} the liquid; in Fig.~\ref{Modifiedsetup}b the CS is located just \emph{outside} the liquid. In the first case, the capillary force can be evaluated by integrating the Young-Laplace pressure over the control surfaces; in the second case, there is no normal force working on these surfaces, but the surface tension acts parallel on the edge of the liquid domain. Both points of view of course result into the same final equations and the resultant forces are indicated by the arrows in Fig.~\ref{Modifiedsetup}b. The resulting momentum balance for the ${\bf x}$ direction reads

\begin{equation}
\rho U^2h\left(\cos\alpha-1\right)=F_{d,x}+F_{men,x}+F_{top,x},
\label{Forbalx}
\end{equation}
while for the ${\bf y}$ direction we have:

\begin{equation}
-\rho U^2h\sin\alpha=F_{d,y}+F_{men,y}+F_{top,y}.
\label{Forbaly}
\end{equation}
Here we used that the Control Volume (CV) contains the full bending of the stream, such that the left and right boundary of the CV are located where the flow is uniform. 

It is instructive to consider the momentum balance in the context of the exact solutions by Keller~\&~Vanden-Broeck~\cite{Keller,Vanden-Broeck1986,Vanden-Broeck1989}. These are obtained by treating the solid as a perfectly sharp edge in the absence of capillary effects. This effectively corresponds to $\We = \infty$, $\tilde{r}_i=0$, while the angles $\theta_0$ and $\beta$ are not defined in this limit. It was found that analytical solutions exist for each value of $\alpha$; one could use the momentum balance (\ref{Forbalx},\ref{Forbaly}) without the capillary forces ($F_{men}=F_{top}=0$) to determine the dynamical force $F_d$, for each value of $\alpha$. Importantly, however, this analysis does not lead to a selection of $\alpha$. A selection of $\alpha$ does appear at finite values of $\We$ and $\tilde{r}_i$, one thus needs explicit expressions for all terms in Sys.~(\ref{Forbalx},\ref{Forbaly}), as will be developed below. In Sec.~\ref{subsec:Calculating local pressures}, we investigate the velocity profile and the pressure distribution in the liquid, from which we compute the various forces in Sec.~\ref{subsec:Calculating the forces}. The resulting set of equations will be presented in Sec.~\ref{subsec:Resulting system of equations}.

\subsection{Flow profiles and pressures} \label{subsec:Calculating local pressures}


To compute the pressure distribution in the liquid we require the velocity profile inside the flowing liquid, for a given geometry characterized by $\tilde{r}_i$. Here, we focus on the case $\tilde{r}_i\gg1$ for which the flow will evolve towards concentric `circular' streamlines around the circular edge of the solid~\cite{Lhuissier}. Under the assumption of potential flow, this corresponds to a free vortex with a tangential flow velocity $u \sim1/r$. Such a profile is quite different from the corner solutions for $\tilde{r}_i\ll1$, and the resulting pressure distribution is expected to be quite different. 

\begin{figure*}[htp!]
\centering
\includegraphics[width=13.0cm]{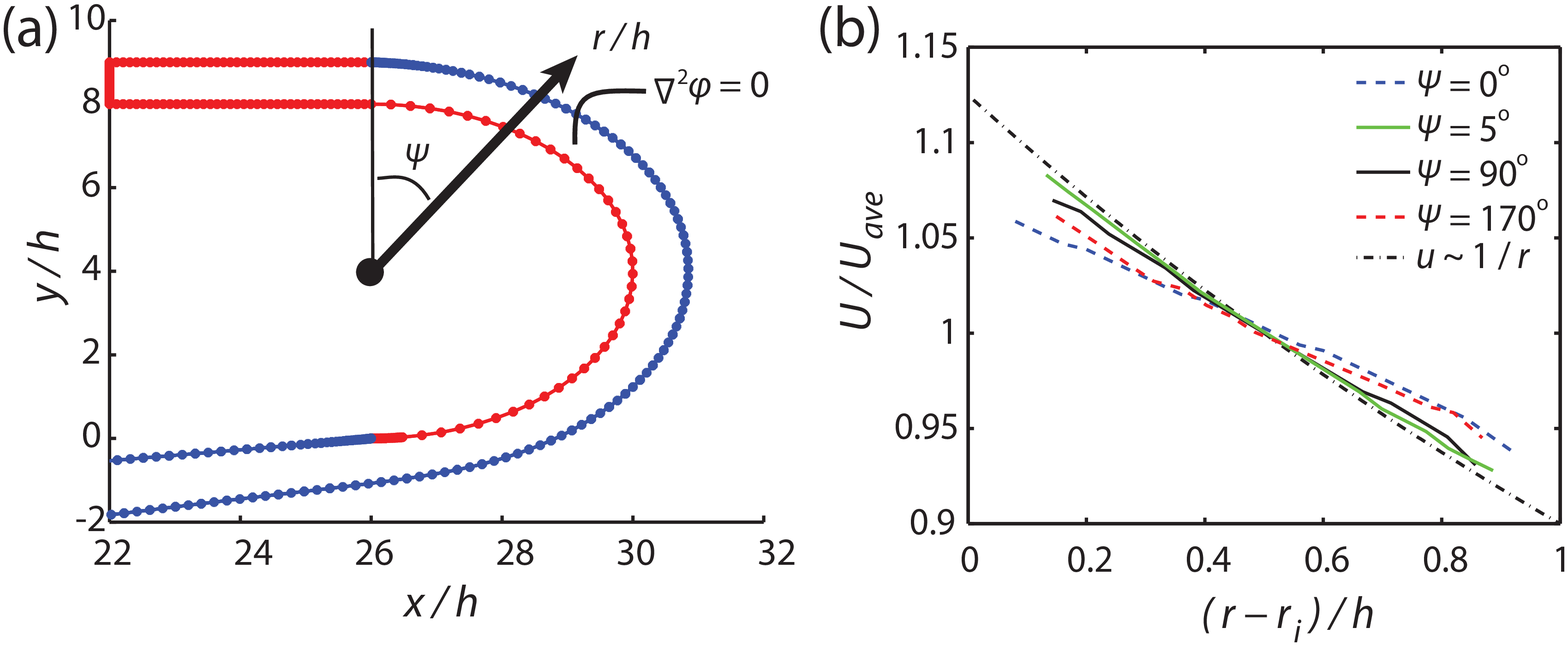}
\caption{Potential flow simulation of the velocity profile in the flowing layer. (a) Sketch of the simulation domain. The red lines are undeformable solid boundaries, while blue edges are deformable and exhibit capillary pressure. The dots are nodes of the boundary integral simulations. Parameters are taken $\tilde{r}_i=4, \We=55$. The simulated setting is in fact axisymmetric, like the experiment using an impacter of radius $R_{imp}$~\cite{Duez}; here we adapted a typical experimental value $R_{imp}/h=26$. (b) Velocity profiles (normalized by the average $U_{ave}$), measured across the liquid film for different locations around the curved solid. The velocity profile quickly evolves towards the expected $1/r$ profile (dash-dotted line).  
}
\label{Streamprofiles}
\end{figure*}

We therefore verified the $1/r$ profiles for an experimentally relevant case, $\tilde{r}_i=4, \We=55$, using potential flow simulations using an axisymmetric Boundary Integral (BI) routine~\cite{Oguz,Bergmann,Gekle,Bouwhuis2013}. We solve the Laplace equation $\nabla^2 \varphi=0$ for the flow potential $\varphi$ in the domain indicated in Fig.~\ref{Streamprofiles}a, containing a small inlet region before the circular bend, ending with a separated sheet. In the simulations the contact line is pinned at a fixed position angle $\beta$. Note that the BI simulations are only used for the confirmation of the presumed velocity profile -- we have not succeeded in creating perfectly steady sheets in the simulations, except for trivial solutions.

The result is shown in Fig.~\ref{Streamprofiles}b, where we plot the normalized velocity profile across the film for various locations $\psi$ along the curved edge (defined in Fig.~\ref{Streamprofiles}a). The initially uniform inflow indeed rapidly evolves towards a $1/r$ profile, indicated as the dash-dotted line.  The entrance/exit effects are quite small: for the example of Fig.~\ref{Streamprofiles} it is not more than a few degrees at the beginning of the bend, and about 10 degrees at the bottom part where the liquid separates from the solid. This shows that a $1/r$ profile for $\tilde{r}_i\gg1$ is a valid approximation. 

Thus, we base our further calculations on the velocity profile $u=A/r$, were the constant $A$ is determined from mass conservation,

\begin{equation}
Uh = \int_{r_i}^{r_i+h} u dr = \int_{r_i}^{r_i+h} \frac{A}{r} dr,
\label{massconservation}
\end{equation}
so that 
\begin{equation}
u=U\left(\frac{1}{\frac{r}{h} \ln(1+\frac{1}{\tilde{r}_i})}\right).
\label{velprof}
\end{equation}
The flow velocity directly above the solid surface is larger than at the top of the liquid sheet, so the pressure difference $P_A-P_B$ (cf. Fig.~\ref{Modifiedsetup}a) is indeed positive from the steady Bernoulli equation:

\begin{eqnarray}
\Delta P&=&P_A-P_B=\frac{1}{2}\rho \left(u|_{r=r_i}^2-u|_{r=r_i+h}^2\right) \nonumber \\
&=&\rho U^2 \left[\frac{\frac{1}{2}+\tilde{r}_i}{\tilde{r}_i^2\left(1+\tilde{r}_i\right)^2\left(\ln(1+\frac{1}{\tilde{r}_i})\right)^2}\right].
\label{Bernoulli}
\end{eqnarray}
Computing $P_A$ (gauge) using the Young-Laplace pressure difference as

\begin{equation}
P_A=\frac{\gamma}{h}\left[\frac{1}{1+\tilde{r}_i}\right],
\label{PA}
\end{equation}
we obtain

\begin{equation}
P_B=\frac{\gamma}{h}\left[\frac{1}{1+\tilde{r}_i}\right]-\rho U^2\left[\frac{\mathcal{G}(\tilde{r}_i)}{\tilde{r}_i}\right].
\label{PB}
\end{equation}
Here we introduced a dimensionless geometrical factor

\begin{equation}
\mathcal{G}(\tilde{r}_i)=\frac{\frac{1}{2}+\tilde{r}_i}{\tilde{r}_i\left(1+\tilde{r}_i\right)^2\left(\ln(1+\frac{1}{\tilde{r}_i})\right)^2},
\label{F}
\end{equation}
which encodes all relevant information of the velocity profile.

Using that $P_B=-\gamma/r_m$, we also extract an expression for $r_m$:

\begin{equation}
\frac{r_m}{h}=\frac{1+\tilde{r}_i}{\We\left(1+\frac{1}{\tilde{r}_i}\right)\mathcal{G}(\tilde{r}_i)-1}.
\label{rm}
\end{equation}
Thus, the size of the meniscus $r_m$ adapts to accommodate the low pressure $P_B$. Note that $\We$ is typically large, and hence $r_m\ll r_i$.

\subsection{Calculating the forces} \label{subsec:Calculating the forces}

We can now proceed to the evaluation of the several terms in the momentum balances~(\ref{Forbalx},\ref{Forbaly}) by integrating the local pressures along the different sections of the CS of Fig.~\ref{Modifiedsetup}a, which is located just inside the liquid. Along the solid, we find for the force induced by the pressure difference projected in ${\bf x}$ direction, $F_{d,x}$:

\begin{eqnarray}
F_{d,x} = \left[\int{P_Bd\mathcal{A}_{wet}}\right]_x = \int_0^{\beta}{P_Br_i\sin\psi d\psi} \nonumber \\
=\left[\rho U^2h\mathcal{G}(\tilde{r}_i)-\gamma\left(\frac{\tilde{r}_i}{1+\tilde{r}_i}\right)\right]\left(\cos\beta-1\right),
\label{FdeltaPx}
\end{eqnarray}
where the integration variable $\psi$ is the angle with respect to the vertical  (see Fig.~\ref{Streamprofiles}a). 
Similarly, we find for $F_{d,y}$:

\begin{eqnarray}
F_{d,y} = \left[\int{P_Bd\mathcal{A}_{wet}}\right]_y = -\int_0^{\beta}{P_Br_i\cos\psi d\psi} \nonumber \\
=-\left[\rho U^2h\mathcal{G}(\tilde{r}_i)-\gamma\left(\frac{\tilde{r}_i}{1+\tilde{r}_i}\right)\right]\sin\beta.
\label{FdeltaPy}
\end{eqnarray}

The capillary forces are induced by the Young-Laplace pressures over the free surface of the meniscus ($F_{men,x}$ and $F_{men,y}$), and over the top free surface of the film ($F_{top,x}$ and $F_{top,y}$). For the capillary force induced by the curvature of the top of the sheet, projected in the $\bf{x}$ direction, we find

\begin{eqnarray}
F_{top,x} &=& -\int_0^{\alpha}P_A\left(r_i+h\right)\sin\psi d\psi \nonumber \\
&=& -\int_0^{\alpha} \left(\frac{\gamma}{r_i+h}\right)\left(r_i+h\right)\sin\psi d\psi \nonumber \\
&=&\gamma\left[\cos\alpha-1\right],
\label{Ftopx}
\end{eqnarray}
and

\begin{equation}
F_{top,y} = -\gamma\sin\alpha.
\label{Ftopy}
\end{equation}
Note that the expressions for $F_{top,x}$ and $F_{top,y}$ are more easily interpreted by considering the explained equivalent picture of the force balance in Fig.~\ref{Modifiedsetup}b: the separate terms are the $\bf{x}$ and $\bf{y}$ projections of the capillary forces along the free surfaces of the CV. A similar observation holds for the capillary forces in the meniscus, which can be obtained by integrating the pressure $P_B$ over the meniscus circle with radius of curvature $r_m$

\begin{eqnarray}
F_{men,x} &=& \int_{-(\pi-\beta-\theta_0)}^{\alpha}P_Br_m\sin\psi d\psi \nonumber \\
&=& \int_{-(\pi-\beta-\theta_0)}^{\alpha} \left(-\frac{\gamma}{r_m}\right)r_m\sin\psi d\psi \nonumber \\
&=&-\gamma\left[\cos(-(\pi-\beta-\theta_0))-\cos\alpha\right] \nonumber \\
&=&\gamma\left[\cos(\beta+\theta_0)+\cos\alpha\right],
\label{Fmenx}
\end{eqnarray}
and 

\begin{equation}
F_{men,y} = \gamma\left[-\sin(\beta+\theta_0)-\sin\alpha\right].
\label{Fmeny}
\end{equation}
Here we used the fact that the meniscus angle, $*$ in Fig.~\ref{Modifiedsetup}, is equal to $\pi-\theta_0-\beta+\alpha$, as follows from the geometry.

\subsection{Resulting system of equations} \label{subsec:Resulting system of equations}

The momentum balance~(\ref{Forbalx},\ref{Forbaly}) combined with the computed forces finally yield the key equations of the model:

\begin{widetext}
\begin{eqnarray}
\left[\We-\We\mathcal{G}(\tilde{r}_i)-\frac{1}{\left(1+\tilde{r}_i\right)}\right] + \left[2-\We\right]\cos\alpha &+& \left[\We\mathcal{G}(\tilde{r}_i)-\frac{\tilde{r}_i}{\left(1+\tilde{r}_i\right)}\right]\cos\beta 
+ \cos(\beta+\theta_0) = 0; \label{xcomp} \\
\left[2-\We\right]\sin\alpha &+& \left[\We\mathcal{G}(\tilde{r}_i)-\frac{\tilde{r}_i}{\left(1+\tilde{r}_i\right)}\right]\sin\beta + \sin(\beta+\theta_0) = 0, \label{ycomp}
\end{eqnarray}
\end{widetext}
with $\We=\rho U^2h/\gamma$, as previously defined. This system should be seen as equations for the angles $\alpha$ and $\beta$, which can be solved for given values of $\We$, $\tilde{r}_i$, and $\theta_0$. Note that by imposing a contact angle $\theta_0$, we by definition consider only separated sheets.

Before we proceed to the numerical solutions of~(\ref{xcomp},\ref{ycomp}) in Sec.~\ref{sec:Results}, we can already analyze a few interesting limits. The limit $\We\rightarrow\infty$ gives as only solution $\alpha=\beta=0$, corresponding to the perfect horizontal sheet, for every value of $\theta_0$. Interestingly, the same holds for $\theta_0=180^o$, for which there is no capillary adhesion: if the surface is superhydrophobic, the jet/sheet will be perfectly horizontal for any $\We$. 
A third interesting case is $\tilde{r}_i\rightarrow\infty$, for which $\mathcal{G}(\tilde{r}_i)\rightarrow1$ and no separated solution exists, unless $\theta_0=180^o$. These trends all agree with the experiments in Ref.~\cite{Duez}. 
A final special case is $\We=2$. In that case, the $\alpha$ dependence completely drops out of the equations, and we are left with two equations for a single unknown, $\beta$. This has no solution unless $\theta_0=180^o$. We can interpret $\We=2$ as a minimum flow speed needed for a non-retracting sheet, as this indeed coincides with the Taylor-Culick velocity~\cite{Taylor,Culick}.

\section{Results} \label{sec:Results}

\subsection{Solutions}

\begin{figure}[htb!]
\centering
\includegraphics[width=8cm]{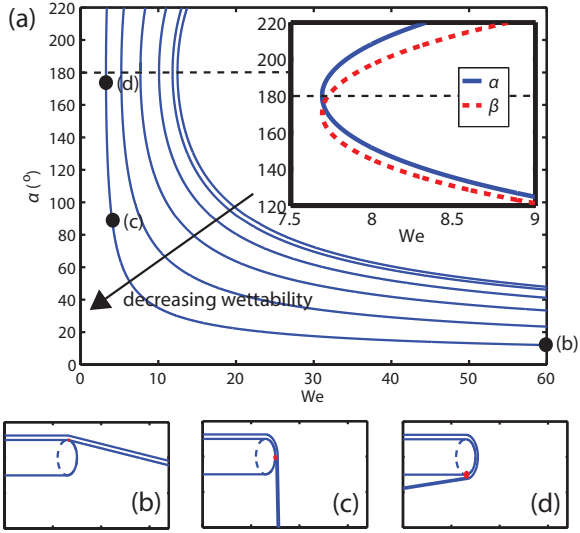}
\caption{(a) $\alpha$ vs $\We=\rho U^2h/\gamma$ for $\tilde{r}_i=4$ and different contact angles. Following the arrow, $\theta_0$ increase from $0^o$ to $150^o$ in steps of $30^o$. Solutions only exist for $\We$ larger than a critical Weber number $\We_c$. Inset: a detailed view of $\alpha$ and $\beta$ as a function of $\We$ around $\We_c$ for $\theta_0=90^o$. (b-d) Resulting flow contours, corresponding to the marked dots in panel (a), showing the dependence of the separation angle on the Weber number for $\theta_0=150^o$.} 
\label{alphavsWe}
\end{figure}

We now analyze the solutions of the momentum balance~(\ref{xcomp},\ref{ycomp}). In Fig.~\ref{alphavsWe}a we report the separation angle $\alpha$ as a function of $\We$ for $\tilde{r}_i=4$. The various curves correspond to different wettabilities, with $\theta_0$ increasing along the arrow. It is found that solutions only exist above a critical value of the Weber number, $\We_c$, which we identify as the threshold for the trickling transition. The critical point is found to coincide with $\alpha=180^\circ$. Above $\We_c$, the momentum balance admits two possible solutions. However, solutions for $\alpha$ larger than $180^o$ are not physical in the sense that the liquid would cross the solid and we focus on the lower solution branch. As expected, the deflection angle $\alpha$ increases when the fluid's inertia is reduced, i.e. as the Weber number is decreased. The sketches in Fig.~\ref{alphavsWe}b-d further illustrate this effect. It should be emphasized that the deflection angle varies rapidly with We for $\alpha$ beyond 90$^\circ$, i.e. when the critical point is approached. The inset of Fig.~\ref{alphavsWe}a shows a zoom around the critical point for both angles, $\alpha$ (blue solid line) and $\beta$ (red dashed line) for $\theta_0=90^o$. The two angles always take similar values, with a maximum difference of about $20^o$. The global minimum of $\beta$ is also reached at $\We_c$, but has a value slightly below $180^o$ (equal to $180^o$ if $\theta_0=0^o$).

\begin{figure}
\centering
\includegraphics[width=8cm]{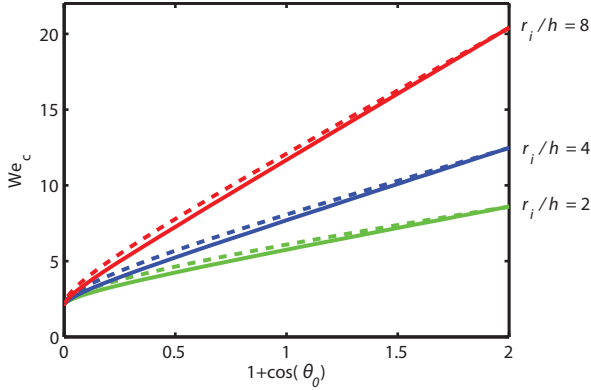}
\caption{Critical Weber number $\We_c$ as a function of $1+\cos{\theta_0}$ for $r_i$ = 0.5, 1, and $2~\mathrm{mm}$, and $h=0.25~\mathrm{mm}$. Solid lines are numerical solutions of the momentum balance; dashed lines correspond to the approximation~(\ref{expansion}).}
\label{Wecagainstwett}
\end{figure}

\begin{figure}[hb!]
\centering
\includegraphics[width=8cm]{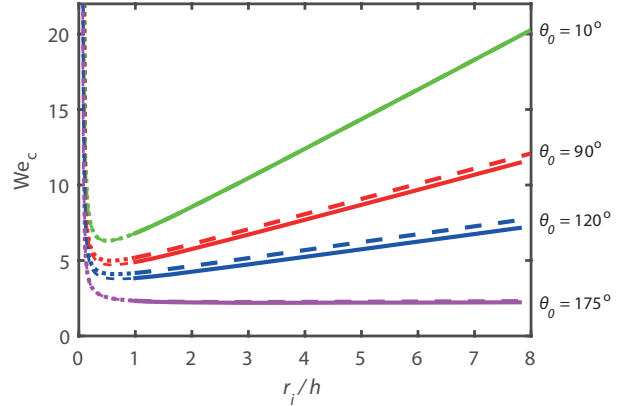}
\caption{Critical Weber number $\We_c$ as a function of $\tilde{r}_i=r_i/h$, $h=0.25~\mathrm{mm}$, for $\theta_0$ = $10^o$, $90^o$, $120^o$, and $175^o$. Solid lines are numerical solutions of the momentum balance; dashed lines correspond to the approximation~(\ref{expansion}). The model is not valid below $\tilde{r}_i\!\sim\!1$, where the predictions are given by a dotted line.}
\label{Wecagainstri}
\end{figure}

The critical Weber number is thus found to depend on the wettability of the solid. Figure~\ref{Wecagainstwett} presents $\We_c$ as a function $1+\cos\theta_0$ for several values of $\tilde{r}_i$ (solid lines). The dependence is approximately linear, with a slope that increases for increasing $\tilde{r}_i$. The linearity no longer applies when $1+\cos\theta_0$ is small, i.e. the superhydrophobic limit, for which all curves converge to $\We_c=2$. In this case there is no adhesion by the solid and the minimal velocity is given by the Taylor-Culick velocity of a free sheet.

Figure~\ref{Wecagainstri} shows $\We_c$ as a function of the radius of curvature of the solid $\tilde{r}_i$, for several values of $\theta_0$ (solid lines). For large $\tilde{r}_i$ we again find a linear trend, reflecting that it is more difficult to separate the sheet when the edge of the solid is not sharp. Note that the model is only valid for $\tilde{r}_i\geq 1$, owing to the assumptions of the velocity field that leads to (\ref{F}). It is still interesting to report the model prediction for $\tilde{r}_i<1$: there is even a divergence that appears when ${\cal G}$ takes the value 1 (Fig.~\ref{Wecagainstri}, dotted lines). In the Discussion section we briefly comment on the limit of small $\tilde{r}_i$. 


\subsection{Asymptotic expansion}

It is instructive to attempt an approximate solution for the critical Weber number, based on the observation that at the critical point $\alpha=180^o$ and $\beta=180^o-\epsilon$, where $\epsilon$ turns out to be small ($<20^o$). Hence, we expand Sys.~(\ref{xcomp},\ref{ycomp}) around up to first order in $\epsilon$, replacing $\sin\epsilon\!\sim\!\epsilon$ and $\cos\epsilon\!\sim\!1$. Using $\sin\left(\theta_0-\epsilon\right)=\sin\theta_0 \cos\epsilon-\cos\theta_0 \sin\epsilon$ and $\cos\left(\theta_0-\epsilon\right)=\cos\theta_0 \cos\epsilon-\sin\theta_0\sin\epsilon$, we find for $\epsilon$ (in radians):

\begin{equation}
\epsilon \approx \frac{\sin\theta_0}{\mathcal{G}(\tilde{r}_i)\We+\cos\theta_0-\frac{\tilde{r}_i}{1+\tilde{r}_i}},
\end{equation}
and

\begin{equation}
\We_c^2 + A\We_c + B \approx 0,
\label{expansion}
\end{equation}
with

\begin{widetext}
\begin{equation}
A = \frac{2\left(\frac{\tilde{r}_i}{1+\tilde{r}_i}\right)+\cos\theta_0\left(3\mathcal{G}(\tilde{r}_i)-2\right)+2\mathcal{G}(\tilde{r}_i)\left(1-\frac{\tilde{r}_i}{1+\tilde{r}_i}\right)+\mathcal{G}(\tilde{r}_i)\left(\frac{1-\tilde{r}_i}{1+\tilde{r}_i}\right)}{-2\mathcal{G}(\tilde{r}_i)+2\mathcal{G}(\tilde{r}_i)^2},
\end{equation}
and

\begin{equation}
B = \frac{-2\left(\frac{\tilde{r}_i}{1+\tilde{r}_i}\right)+2\cos\theta_0+\frac{\tilde{r}_i\left(\tilde{r}_i-1\right)}{\left(1+\tilde{r}_i\right)^2}-\left(\frac{2\tilde{r}_i-1}{1+\tilde{r}_i}\right)\cos\theta_0+1}{-2\mathcal{G}(\tilde{r}_i)+2\mathcal{G}(\tilde{r}_i)^2}.
\end{equation}
\end{widetext}
This gives a quadratic equation for $\We_c$, which is superimposed as dashed lines in Figs.~\ref{Wecagainstwett} and \ref{Wecagainstri}. Indeed, this approximate solution gives a very good description of the full solutions. The difference is largest for $\theta_0\!\sim\!90^o$, where the largest values for $\epsilon$ are encountered.

We could even further simplify (\ref{expansion}) when $\tilde{r}_i\!\gg\!1$. $\mathcal{G}(\tilde{r}_i)$ can then be approximated by $1\!-\!1/\left(2\tilde{r}_i\right)$), and this yields

\begin{equation}
\We_c \sim \tilde{r}_i\left(1+\cos\theta_0\right).
\label{Wecexp}
\end{equation}
This predicts linear behavior of the critical Weber number with respect to both the aspect ratio $\tilde{r}_i$ and to $1+\cos\theta_0$, which is consistent with our results in Figs.~\ref{Wecagainstwett} and \ref{Wecagainstri}.

\section{Discussion}\label{sec:Discussion}

\begin{figure*}
\centering
\includegraphics[width=15.0cm]{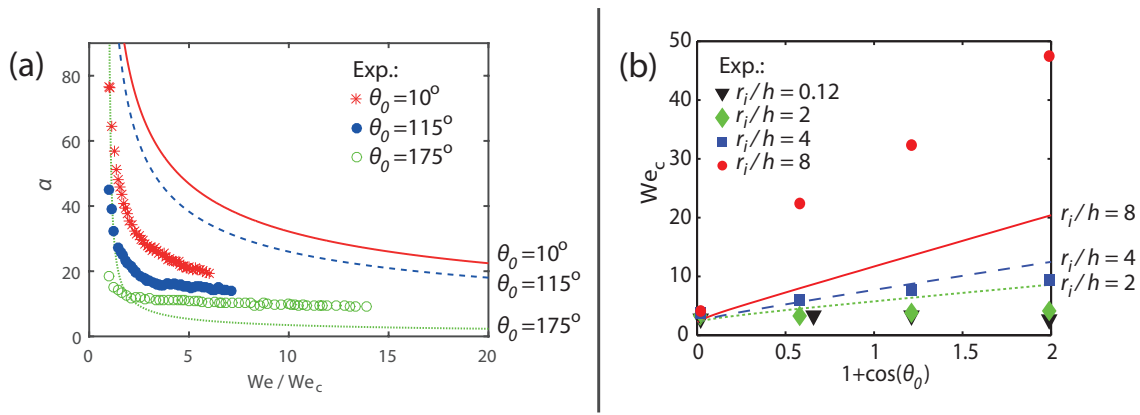}
\caption{Quantitative comparison between our model and experimental data taken from Duez~{\it et~al.}~\cite{Duez}. (a) $\alpha$ vs. $\We/\We_c$ for $\theta_0$ = $10^o$, $115^o$, and $175^o$; $r_i=1~\mathrm{mm}$ and $h=0.25~\mathrm{mm}$; (b) $\We_c$ vs. $(1+\cos\theta_0)$ for several $\tilde{r}_i=r_i/h$.}
\label{expcomp}
\end{figure*}

We have performed a control volume analysis of inertial pouring flows, taking into account the fluid inertia, surface tension, and the curvature of the edge of the solid. The analysis explicitly recovers that steady states can only exist above a critical Weber number $\We_c$, and thus captures the transition to trickling. The work also recovers the experimental trends that $\We_c$ decreases as the solid is sharper and more hydrophobic, and identify a lower bound $\We_c\geq 2$. Here we wish to conclude by making a direct comparison to experiments~\citep{Duez,Dong}.

Figure~\ref{expcomp}a reports the separation angle $\alpha$ against $\We/\We_c$ for three different contact angles, $\theta_0=10^o$, $115^o$, and $175^o$. The value of $\tilde{r}_i=4$ was adapted from the experiment, and the symbols represent data from Ref.~\cite{Duez}. It is clear that the model captures the experimental trends, but is limited in terms of quantitative prediction. In particular, the change of angle $\alpha$ close to the critical point is underpredicted by the model. Also, the experiments at large Weber number exhibit a sheet deflection of the order of 10$^\circ$, which points to either dissipation in the fluid or an influence of the axisymmetric setup -- both of which are not taken into account in the model.  

In Fig.~\ref{expcomp}b, we show the critical Weber number $\We_c$ as a function of $1+\cos\theta$, for three different $\tilde{r}_i$. The model nicely captures the linear dependence $\We_c\sim (1+\cos\theta_0)$ that is observed in experiment. In addition, the data in Fig.~\ref{expcomp}b are consistent with the predicted lower bound on the critical Weber number of $\We_c=2$, set by the Taylor-Culick velocity. The model underestimates the dependence on $\tilde{r}_i$: we predict a linear increase of $\We_c$ with $\tilde{r}_i$, while the experimental data are better described by $\tilde{r}_i^2$, as was also proposed by a scaling argument~\cite{Duez}. However, the recent data in Ref.~\cite{Dong} are best described by a linear dependence on $\tilde{r}_i$, even though their experimental setup is in principle comparable. At present, there is thus some uncertainty on how the sharpness of the edge affects the trickling transition. We do note that the model predictions shown here are without adjustable parameters and the discrepancy with, and between, various experiments is always less than a factor of two. 

The momentum balance presented in this paper has the merit that it provides a general framework for future investigations of pouring flows. Namely, the capillary forces are described correctly here -- apart perhaps from the question of whether $\theta_0$ can be interpreted as the equilibrium contact angle~\citep{Bonn,Snoeijer} -- and the main assumption is on the estimate of the hydrodynamic force $F_d$. As mentioned already, we only estimated this force in the case of large $\tilde{r}_i$, but (\ref{Forbalx},\ref{Forbaly}) can in fact be used to estimate $F_d$ based on experimental data. The black triangles in Fig. \ref{expcomp}b correspond to experiments with $\tilde{r}_i=0.12$, for which $\We_c$ exhibits almost no dependence on wettability and is close to the lower bound $\We_c=2$. The momentum balance suggests that $F_d \rightarrow 0$ in this limit, implying a vanishing hydrodynamic retention around sharp edges. Interestingly, this is a strong departure from the exact potential flow solutions around a perfectly sharp edge~\cite{Vanden-Broeck1986,Vanden-Broeck1989}, which do not include capillary adhesion, for which trickling completely relies on a nonzero $F_d$. Future work should further reveal how the presence of a meniscus on a sharp edge influences trickling in inertial pouring flows.

\acknowledgments

We are grateful to Christophe Clanet for pointing out the relation with the Taylor-Culick velocity. We also thank Sander Wildeman and Henri Lhuissier for their helpful suggestions. This work was funded by VIDI Grant. No. 11304, which is financially supported by the `Nederlandse Organisatie voor Wetenschappelijk Onderzoek (NWO).'

\end{document}